\documentclass[aip,graphicx,reprint,floatfix]{revtex4-1}

\usepackage{graphicx}% Include figure files
\usepackage{color}
\usepackage{amsmath}
\usepackage{mathtools}
\usepackage[ngerman,english]{babel}
\usepackage{amsfonts}
\usepackage{amsmath}
\usepackage{array}
\usepackage{epsf}
\usepackage{epsfig}
\usepackage{graphics}
\usepackage{epstopdf}
\usepackage{footnote}
\usepackage{bbm}
\usepackage{braket}

\newcommand{\rt}[1]{\textcolor{red}{#1}} % critical comments
\newcommand{\bt}[1]{\textcolor{black}{#1}} % 
\newcommand{\Fe}{[Fe(H$_2$O)$_{6}$]$^{2+}$}

\newcommand{\Supp}{Supplement}

\makeatletter
\let\l@ENGLISH\l@english
\makeatother

\begin{document}

\title{Ultrafast dissipative spin-state dynamics triggered by  X-ray pulse trains}
\author{Huihui Wang}
\author{Tobias M\"{o}hle}
\author{Oliver K\"{u}hn}
\author{Sergey I. Bokarev}
\email{sergey.bokarev@uni-rostock.de}
\affiliation{ Institut f\"{u}r Physik, Universit\"{a}t Rostock,
  Albert-Einstein-Str. 23-24, 18059 Rostock, Germany}

\begin{abstract}
Frontiers of attosecond science are constantly shifting, thus addressing more and more intricate effects with increasing resolution. Ultrashort pulses offer a practical way to prepare complex superpositions of quantum states, follow, and steer their dynamics. In this contribution, an ultrafast spin-flip process triggered by sub-femtosecond (fs) excitation and  strong spin-orbit coupling between 2p core-excited states of a transition metal complex is investigated using density matrix-based time-dependent restricted active space configuration interaction theory. The effect of the nuclear vibrations is incorporated making use of an electronic system plus vibrational bath partitioning. The differences between isolated sub-fs pulses and pulse trains as well as influence of various pulse characteristics on the initiated dynamics are discussed.  The effect under study can be potentially used for ultrafast clocking in sub-few fs experiments.
\end{abstract}

  
\maketitle

%--------------------------------------------------------------
\section{Introduction}
\label{sec:intro}
%--------------------------------------------------------------

%\begin{itemize}
% \item attosecond experiments
% \item addressing ultimate timescales of electronic motion
% \item first step -- preparation of superpositions of states and then following subsequent time evolution
% \item previous works -- electron dynamics for the case of strong SOC
% \item peculiarity is that nuclei were fixed and the dynamics were triggered by an isolated pulse (which require additional effort to obtain)
% \item Here we try to include influence of vibrations at the level of system (electronic degrees of freedom)/bath (nuclear degrees of freedom) partitioning and consider the dynamics caused by trains of pulses
%\end{itemize}

Constant decrease of the duration of light pulses achieved in experiments sharpens the spectroscopic probe used to dissect different atomic and molecular ultrafast processes. 
This applies not only to nuclear dynamics occurring on the timescale of tens to hundreds of femtoseconds but also to electron dynamics happening about three orders of magnitude faster.~\cite{Schultz_book_2014,Lepine_CPL_2013}
In particular, the unprecedented insight into the electron's ``chemistry'' has become possible due to development of free-electron lasers and high-harmonic generation sources of ultrashort pulses.~\cite{Young_JPB_2018}

These pulses have sufficient bandwidth to create a non-stationary superposition of several electronic states and to follow the dynamics of the initiated wavepacket which can be caused both by nuclear and electronic motion. Although experimental developments are only at the beginning of their way, bright prospects are seen and the topic is actively developed by theoreticians (for review see, e.g., Refs.~\citenum{Moskalenko_PR_2017,Kuleff_JPB_2014,Zhang_TCC_2015}). For instance, the process of charge migration, where the dynamics of an electronic wavepacket being a superposition of ionized states has been extensively studied theoretically. In this case, the driving force of the dynamics is electron correlation and relaxation. The nuclear motion is argued either to be of minor importance or to cause an ultrafast dephasing destroying coherent electron dynamics. For a discussion of the relevance of  nuclear motion for the early-time electron dynamics, see, e.g., Refs.~\citenum{Mendive-Tapia_JCP_2013, Li_PRL_2015, Vacher_PRA_2015, Despre_JPCL_2015,Arnold_PRA_2017}.
%The particular effect of nuclear motion shall depend on the involvement of conical intersections and the number of thermally accessible conformers. 

Recently, ultrafast electron dynamics driven by strong spin-orbit coupling (SOC) in the 2p core-excited states of an iron complex have been reported.~\cite{Wang_PRL_2017,Wang_MP_2017}  Remarkably, the dynamics occur on a timescale faster than the Auger decay lifetime of few femtoseconds.  These studies considered excitation by a single isolated X-ray pulse, thus, resembling a free electron laser experiments. However, the experimental realization with a High Harmonic Generation (HHG) setup would require additional effort to obtain such isolated pulses if compared to trains of pulses which are more common in this case.~\cite{Schultz_book_2014} Therefore, in the present work, we extend previous studies for the exemplary case of Fe$^{2+}$ in its first solvation shell along two lines: First, we consider the dynamics driven by trains of ultrashort pulses, which are easier to generate in HHG experiments. Second, we examine the decoherence caused by nuclear vibrations. To this end, a  system (electronic degrees of freedom)/bath (nuclear degrees of freedom) model is introduced, whose dynamics is treated using a Quantum Master Equation.   In~Sec.~\ref{sec:theory}, the general framework of the employed time-dependent formalism is presented, followed by the description of computational details in Sec.~\ref{sec:comp}.  The consequences of dissipation caused by the vibrational bath as well as dynamics caused by series of pulses are discussed in Sec.~\ref{sec:results}. Conclusions and outlook are given in Sec.~\ref{sec:concl}.
%
%--------------------------------------------------------------
\section{Theory}
\label{sec:theory}
%--------------------------------------------------------------

In the following, we make use of the Born-Oppenheimer approximation, thus assuming that the system is excited far from conical intersections.
%(For a discussion of the relevance of  nuclear motion for the early-time electron dynamics, see also Refs.~\citenum{Mendive-Tapia_JCP_2013, Li_PRL_2015, Vacher_PRA_2015, Despre_JPCL_2015}.) 
The actual dynamics are studied at a single geometry of the molecule, assuming vertical excitation by the incoming light. The timescale of the processes discussed here (few femtoseconds) is much shorter than the periods of relevant nuclear vibrations (see discussion in Sec.~\ref{sec:results}), what builds the basis for the present model.

In order to account at least approximately for a possible influence of molecular vibrations but also for loss channels such as Auger decay, a system-bath approach will be used, i.e. the total Hamiltonian is written as
\begin{equation}\label{eq:TotHam}
\hat{H}(t)=\hat{H}_{\rm  S}(t)+\hat{H}_{\rm B}+\hat{H}_{\rm S-B} \,.
\end{equation}  
Here, $\hat{H}_{\rm S}(t)$ describes the relevant system, i.e. that part of the electronic degrees of freedom (DOF) whose dynamics is triggered by the X-ray light.  The relevant system is coupled via $\hat{H}_{\rm S-B} $ to some bath with Hamiltonian $\hat{H}_{\rm B}$. The dynamics of the relevant system according to the reduced density operator, $\hat\rho$, follows from the Quantum Master Equation ($\hbar=1$)~\cite{May_book_2011}
 \begin{equation}\label{eq:LvN}
    \frac{\partial}{\partial{t}}\hat{\rho}=-i[\hat{H}_{\rm S}(t),\hat{\rho}]+\mathcal{R}\hat\rho\ +\mathcal{A}\hat\rho \,.
  \end{equation}
  Equation \eqref{eq:LvN} assumes that the effect of the system-bath interaction can be treated in second order perturbation theory and invoking the Markov approximation. Here, $\mathcal{R}$ is the dissipation superoperator, which accounts for phase and energy  relaxation due to interaction with the vibrational bath and $\mathcal{A}$ accounts for the Auger decay.
The framework for the description of the electronic subsystem with the density matrix-based restricted active space configuration interaction ($\rho$-TD-RASCI) is introduced in Section~\ref{sec:tdci} and the construction of $\mathcal{R}$ and $\mathcal{A}$ is described in Section~\ref{sec:dissipation}.

%, which contain two parts $H_{CI}$ and $V_{SOC}$, the former is the configuration interaction (CI) Hamiltonian representing the effect of the electron correlation, while the latter parts $V_{SOC}$ is the spin-orbit coupling which makes use of the atomic mean-field integral approximation and effective one-electron approximation to the Breit-Pauli equation within LS-coupling scheme. 
%In the present case of the interaction of a molecular system with X-ray light, where electronic transitions possess a very local character, the focus is put on a fairly small subsystem containing the absorbing atom with its first coordination shell.
%To account for dissipation due to the more extended environment or electronic relaxation processes, which are not treated explicitly, it is natural to represent the system in terms of the evolving in time according to
%
%The question whether the nuclear motion can be neglected for the early-time dynamics has triggered an ongoing debate~\cite{Mendive-tapia_2013,Li2015, Vacher_2015, Despre_2015}. Here, we assume that the system is excited far from conical intersections and that the considered time interval is shorter than the relevant vibrational periods.
 
%The dynamics of the system will be described in terms of the reduced density matrix (RDM), $\hat{\rho}$, taking the trace over the vibrational degrees of freedom according to the following equation of motion:

%--------------------------------------------------------------
\subsection{$\rho$-TD-RASCI}
\label{sec:tdci}
%--------------------------------------------------------------

%Ultrafast spin-flip is investigated using the Time-Dependent Restricted Active Space Configuration Interaction method in its  density matrix formulation ($\rho$-TD-RASCI), which is similar in spirit to the techniques proposed in Refs.~\citenum{Tremblay2011,Kato2012,Jin_2012}. 
%As compared to TD-RASCI, the density matrix formulation offers a convenient way of treating dissipative dynamics of open quantum systems.
%
%Working within Born-Oppenheimer approximation and provided that processes under study occur much faster than the period of nuclear motion we assume the clamped nuclei approximation. Here, the nuclei are fixed at the ground state equilibrium positions, and thus we solve the electronic \Sch equation only. 

Within the $\rho$-TD-RASCI method, described elsewhere in detail,\cite{Wang_PRL_2017, Wang_MP_2017} see also Refs.~\citenum{Tremblay_JCP_2011,Kato_PTPS_2012,Jin_PRL_2012}, the reduced density operator is represented in the basis of electronic Configuration State Functions (CSFs), $\{ \Phi_j^{(S,M_S)}\}$, with the total spin $S$ and its projection $M_S$. Note that the relaxation of one-electron molecular orbitals (MO) is not taken into account and CSFs are constructed using a time-independent MO basis, optimized at the restricted active space self-consistent field~\cite{Malmqvist_JPC_1990} level prior to propagation. On this   level, $\rho$-TD-RASCI can describe electron correlation-driven processes (see regime III in Ref.~\citenum{Wang_MP_2017}) analogous to charge migration in ionized species.~\cite{} 

\begin{figure}
\includegraphics[width=0.45\textwidth]{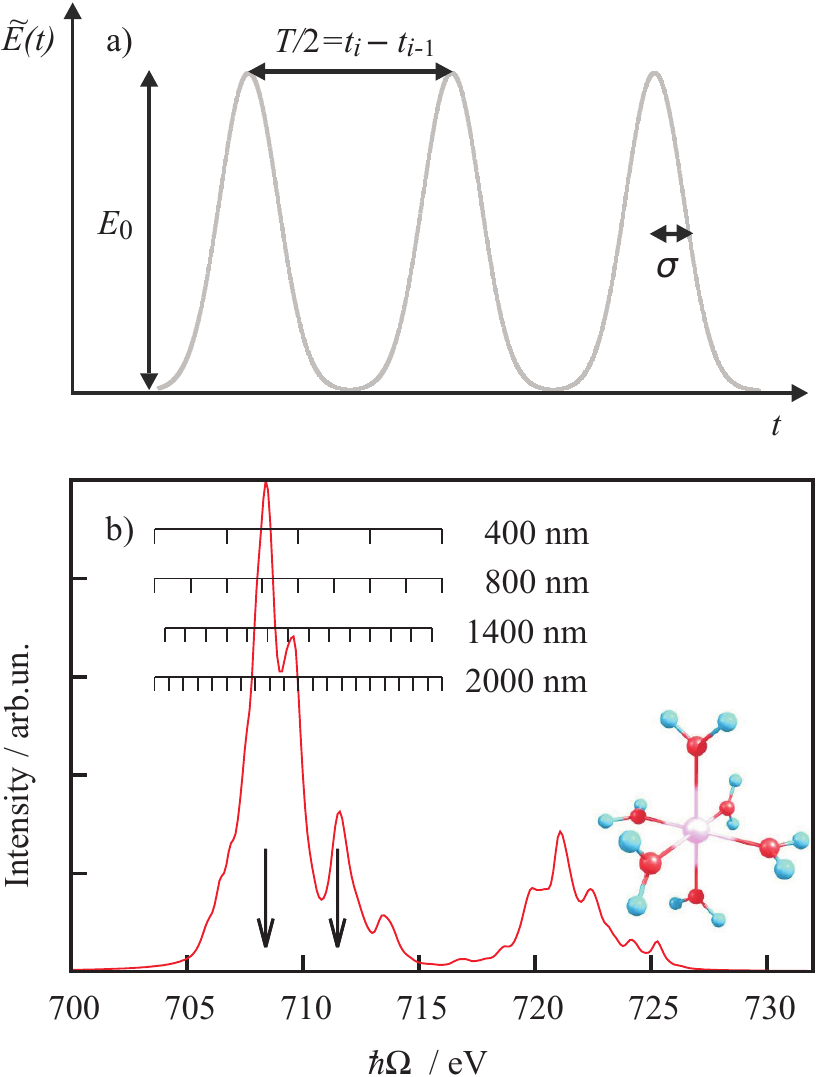}
\caption{a) Scheme explaining the parameters of the light pulse envelope (Eq.~\ref{eq:pulse}); see also Table~\ref{tab:pulse}. $T$ is the period of the driving laser optical cycle. 
b) Calculated X-ray absorption spectrum. The energies corresponding to carrier frequencies $\hbar \Omega=$708.4 and 711.5\,eV are marked with arrows. The frequency combs corresponding to HHG with the respective wavelengths of the driving laser are also shown. The structure of the \Fe complex is presented as an inset.
\label{fig:multi_pulse_scheme}
}
\end{figure}

The Hamiltonian in the CSF basis reads
%
%-------------------------------------
\begin{eqnarray}
	\label{eq:Ham}
   \mathbf{H}_{\rm S}(t)&= &\mathbf{H}_{\rm CI}+\mathbf{V}_{\rm SOC}+\mathbf{U}_{\rm ext}(t) \nonumber \\
   &=&
   \left(
   \begin{array}{cc} 
      \mathbf{H}_{h}  &  0 \\
      0  & \mathbf{H}_{l} \\
   \end{array}
   \right)
   +   
   \left(
   \begin{array}{cc} 
      \mathbf{V}_{hh}  &  \mathbf{V}_{hl} \\
      \mathbf{V}_{lh}  &  \mathbf{V}_{ll} \\
   \end{array}
   \right) 
   +
  \left(
  \begin{array}{cc} 
      \mathbf{U}_{h}(t) &  0 \\
      0  &  \mathbf{U}_{l}(t) \\
   \end{array}
   \right), \nonumber \\
   \,
\end{eqnarray}
%-------------------------------------
where we separated blocks of low- ($l$) and high-spin ($h$) basis functions. In Eq.~\eqref{eq:Ham}, $\mathbf{H}_{\rm CI}$ is the configuration interaction (CI) Hamiltonian matrix containing the effect of electron correlation.  SOC is contained in $\mathbf{V}_{\rm SOC}$, whose matrix elements are calculated in the LS-coupling limit, making use of the atomic mean-field integral approximation.~\cite{Hess_CPL_1996} The interaction with the time-dependent electric field, $\mathbf{U}_{s=h,l}(t)=-\vec{\mathbf{d}}_{ss} \cdot \vec{E}(t)$,
is taken in  semiclassical dipole approximation with the transition dipole matrices $\vec{\mathbf{d}}_{hh}$ and $\vec{\mathbf{d}}_{ll}$ ($\vec{\mathbf{d}}_{hl}=\vec{\mathbf{d}}_{lh}=\vec{\mathbf{0}}$). 
The electric field, $\vec{E}(t)$, corresponds to a train of ultrashort pulses with temporal positions $t_i$, polarizations $\vec{e}$, carrier frequencies $\Omega$ and Gaussian envelopes with width $\sigma$, modeling HHG output, (cf. Fig.~\ref{fig:multi_pulse_scheme}a)): 
\begin{align}\label{eq:pulse} 
\vec{E}(t) &= \vec{e}\tilde{E}(t)\cos(\Omega t) \\
\tilde{E}(t) &= E_0\sum_i\exp(-(t-t_i)^2/(2\sigma^2)). \nonumber
\end{align} 
The delay between two subpulses, $t_i-t_{i-1}$, is determined by the half-period of the optical cycle of the HHG driving laser. The influence of carrier-envelope phase is considered to be irrelevant for the carrier frequency $\Omega$ in X-ray range, since multiple oscillations occur within a single pulse in a train.

The eigenstates of $\mathbf{H}_{\rm CI}$ will be called spin-free (SF) states, they are given by the set $\{\Psi^{{\rm SF},(S,M_S)}_{n}\}$. This set can be used to express the eigenstates of $\mathbf{H}_{\rm CI}+\mathbf{V}_{\rm SOC}$, which are called SOC states, as follows:
\begin{equation}\label{eq:SOC_states}
 \ket{\Psi^{\rm SOC}_{a}}=\sum_{n}C_{an}^{(S,M_S)} \ket{\Psi^{{\rm SF},(S,M_S)}_{n}} \, .
\end{equation}
The notation of the coefficients implies that there is a correspondence $n\leftrightarrow (S,M_S)$.
The SOC basis will be used for representation of the reduced density operator in Eq.~\eqref{eq:LvN}, which takes the form
 \begin{eqnarray}\label{eq:LvN_SOC}
    \frac{\partial}{\partial{t}}\rho_{ab}&=&-i\omega_{ab}\rho_{ab} +i {\vec E}(t) \sum_{c} ({\vec d}_{ac}\rho_{cb} - {\vec d}_{cb}\rho_{ac})\nonumber\\
    &-& \sum_{cd} (\mathcal{R}_{ab,cd}+\mathcal{A}_{ab,cd}) \rho_{cd} \, .
  \end{eqnarray}
Thus, the time-propagation is performed in the SOC basis which ensures the rigorous structure of the respective equations of motion for the dissipative dynamics leading to proper equilibration.\cite{May_book_2011}
However, the analysis below will also involve the populations  of SF states for a certain multiplicity. They are obtained from
\begin{equation}
	\rho_{nn}^{(S)}=\sum_{ab} C_{an}^{(S,M_S)}[C_{bn}^{(S,M_S)}]^* \rho_{ab} \,. 
\end{equation}
%

% % % % % % %
\iffalse
The structure of the SF state can be viewed as linear combination
\begin{equation}\label{wave function}
    |\Psi^{\rm SF}\rangle=C_0 \ket{\Psi_{0}}+\underbrace{\sum_{ia}C_{i}^{a}\ket{\Psi_{i}^{a}}}_{\rm 1h1p}+\underbrace{\sum_{i<j,a<b}C_{ij}^{ab}\ket{\Psi_{ij}^{ab}}}_{\rm 2h2p}+\ldots,
\end{equation}
where $\ket{\Psi_{0}}$ represents the reference wave function of the ground electronic state and the remaining terms are the single (1h1p), double (2h2p), etc excitations from the ground state configuration. 
Indices $i,j,\ldots$ denote occupied MOs from which electron is excited to unoccupied $a,b,\ldots$ ones.
\fi
% % % % % % % %
%Employing the restricted active space strategy one can conveniently choose the basis of CSFs involving excitations from/to orbitals of interest, being systematically improvable up to the exact limit. 
%

%The latter is an effective one-electron approximation to the Breit-Pauli equation, which has demonstrated good performance for  L-edge spectra of transition metal compounds~\cite{Josefsson2012,Bokarev2013,Suljoti_2013,Atak2013,pinjari_2014,engel_2014,Bokarev2015,wernet_2015,Grell2015,pinjari_2016,Golnak2016,Preuse_2016}.  
%It   provides an intuitive interpretation in terms of pure spin-states with well-defined $S$ and $M_S$ quantum numbers. 

%

%--------------------------------------------------------------
\subsection{Dissipation Operators}
\label{sec:dissipation}
%--------------------------------------------------------------
%Alternatively to the harmonic treatment one can study the interstate coupling autocorrelation function in interaction representation to get memory kernels formulae 3.181 and 3.182 on page 106 of Oliver's book.

First, we focus on $\mathcal{R}$ which describes the effect of electron-vibrational coupling.  Having in mind the case of 
a Fe$^{2+}$ ion in its first solvation shell, it is assumed that the vibrations of [Fe(H$_2$O)$_6$]$^{2+}$ can be mapped onto a harmonic oscillator model. In other words, the vibrational bath is assumed to be a collection of harmonic oscillators in thermal equilibrium coupled to the electronic transitions in a Huang-Rhys like fashion.~\cite{May_book_2011} This constitutes the primary vibrational bath with the interaction operator

%Assuming that it has a quite rigid structure, we employ the harmonic oscillator model to describe its vibrations and consider respective energy dissipation within an advanced kinetic model. 
%Note that if we want to calculate the influence of the liquid environment beyond first solvation shell classical simulations of the bath correlation function are required.}
%interaction between the system and bath which is dependent on the system and bath operators. In generally, it can be factored into the arbitrary functions of system and bath operators, here we use Caldeira-Leggett model which only consider the lowest-Taylor expansion with respect to the system function and bath function as following equation:
%\begin{equation}
%H_{s-b}=\sum_{i}\sum_{\zeta}c_{i,\zeta}X_{i}Q_{\zeta}
%\end{equation}
%here, $X_{i}$ represents the bath part--an harmonic oscillator position operator and operator $Q_{\zeta}$ denotes the part of the system. The coefficients $c_{i,\zeta}$ indicates the system-bath coupling.  Note that the linear expansion with respect to the system operators restricts the description to energy gap fluctuations in the system.
%
%The adjustment of the Caldeira-Leggett system-bath Hamiltonian is necessary to adapt different system-bath models. In our models the system couples to the primary bath of the intra-molecular DOFs (vibration oscillator) which couple to secondary baths of environment DOFs. The system-bath Hamiltonian can be described as:
\begin{equation}
\hat{H}_{\rm el-vib}=\sum_{ab}\sum_{\xi}g_{ab,\xi}\omega_{\xi}Q_{\xi}\ket{\Psi^{\rm SOC}_{a}}\bra{\Psi^{\rm SOC}_{b}} \, ,
\end{equation}
where  $a$ and $b$ label the coupled electronic states and $Q_\xi$ is the coordinate of the normal mode $\xi$ having frequency $\omega_{\xi}$. 
Here, $g_{ab,\xi}$ is the dimensionless shift of the $a$'s state harmonic potential energy surface with respect to the potential of state $b$, which can be expressed by the Huang-Rhys factor $S_{ab,\xi}=g_{ab,\xi}^{2}/2$. 
The coupling strengths $g_{ab,\xi}$ in the SOC basis have been obtained by the transformation of factors $g^{\rm SF}_{0n,\xi}$, evaluated in the  basis of SF states with respect to the ground state $n=0$,  according to Eq.~\eqref{eq:SOC_states}. Note that one could further take the non-adiabatic couplings into account explicitly within an approach described in Ref.~\citenum{Hermann_JPCC_2015}.

This primary bath describing [Fe(H$_2$O)$_6$]$^{2+}$ is coupled to a secondary one (further solvation shells) leading to a multi-mode Brownian oscillator model.~\cite{Mukamel_book_1999} 
In total, the effect of primary and secondary bath can be described by the following spectral density

\begin{equation}\label{eq:J}
J_{ab}(\omega)=\sum_{\xi}\omega_{\xi}^{2}g_{ab,\xi}^{2}\frac{\omega\omega_{\xi}\gamma}{(\omega^{2}-\omega_{\xi}^{2})^{2}+\omega^{2}\gamma^{2}}\, ,
\end{equation}
%
%\begin{equation}
%J(\omega)=\pi\sum_{\xi}\omega_{\xi}^{2}g_{\xi}^{2}\delta(\omega-\omega_{\xi})
%\end{equation}
%The bath correlation function can be expressed via the spectral density as
%\begin{equation}
%C(\omega)=2[1+n(\omega_{\xi})][J_{\xi}(\omega)-J_{\xi}(-\omega)]
%\end{equation}
where the parameter $\gamma$ accounts for the influence of the secondary bath.

%, it always be used to characterise the coupling between the electronic and vibrational DOFs. The last part $Q_{\xi}|a\rangle\langle a|$ is the system coordinates.
%As the same spirit of the system-bath hamiltonian, the density operator can also factors into system and bath part, and it is the so called reduced density operator which can simplified the time-dependent Schr\"{o}dinger  equation. Then with the Markov approximation the motion for the reduce density matrix (RDM) in energy representation can be derived as:
%\begin{equation}
%\frac{\partial}{\partial t}\hat{\rho}_{ab}=-i\omega_{ab}\hat{\rho}_{ab}-\frac{1}{\hbar}\sum_{c}(V_{ac}\hat{\rho}_{cb}-V_{cb}\hat{\rho}_{ac})-\sum_{cd}R_{abcd}\hat{\rho}_{cd}
%\end{equation}
%The dissipative contribution to the RDM equations of motion can be finally written as
%\begin{equation}
%(\frac{\partial}{\partial t}\hat{\rho})_{diss}=-\sum_{cd}R_{abcd}\hat{\rho}_{cd}(t)
%\end{equation}
%
For simplicity we will restrict ourselves to the Bloch model which decouples population relaxation and coherence dephasing. 
In this case, the only non-zero elements of the relaxation matrix (Redfield tensor)  ${\mathcal R}_{ab,cd}$ are given by
%
%For population transfer which is the diagonal element, fulfil $a=b, c=d$, the correspond matrix element can be written as
\begin{equation}\label{eq:population}
{\mathcal R}_{aa,cc}=\delta_{ac}\sum_{e} k_{a\rightarrow e}-k_{c\rightarrow a}
\end{equation}
for  population relaxation  and
\begin{equation}\label{eq:dephasing}
{\mathcal R}_{ab,ab}=\frac{1}{2}(\sum_{e}k_{a\rightarrow e}+\sum_{e}k_{b\rightarrow e}) %+\gamma_{ab}^{(pd)}
\end{equation}
for coherence dephasing.
%
%\begin{equation}\label{eq:auger}
%\mathcal{D^{\rm SOC}}=-\sum_k \Gamma_k \ket{\Psi^{\rm SOC}_k} \bra{\Psi^{\rm SOC}_k}.
%\end{equation}
%
%The respective $\mathcal{D^{\rm SOC}}$ matrix is then transformed to the CSF basis and used for the propagation (Eq.~\eqref{eq:LvN}).
%The pure dephasing term $\gamma_{ab}^{(pd)}$ in Eq.~\ref{eq:dephasing} \bt{is set to zero in current implementation.}
%Note that in the form of Eqs.~\ref{eq:population} and \ref{eq:dephasing} (which is also called Lindblad form) 
%the structure of the $R$-tensor ensures probabilistic treatment of the $\rho$ diagonal and conserves total population.
%However, the term due to the Auger decay leads to exponential decay of the trace of $\rho$.
%Look what Saalfrank or some other authors (see refs in proposal) treat pure dephasing.
%\begin{equation}
%\gamma_{ab}^{(pd)}=-\sum_{\mu,\nu}K_{aa}^{(\mu)}K_{bb}^{(\nu)}C_{\mu\nu}(\omega=0)
%\end{equation}
%SIB: K can be just omitted, see section 3.8.4 of Oliver's book
The relaxation rates $k_{a\rightarrow b}$ for the transition from a state $\ket{\Psi^{\rm SOC}_{a}}$  to a state $\ket{\Psi^{\rm SOC}_{b}}$, can be expressed as
\begin{equation}\label{eq:k_ab}
k_{a\rightarrow b}=2[1+n(\omega)][J_{ab}(\omega)-J_{ab}(-\omega)] \, ,
\end{equation}
%
%Here, the expression of the correlation function is
%\begin{equation}
%C(\omega)=2\pi\sum_{\xi}(\omega_{\xi}g_{\xi})^{2}([1+n(\omega_{\xi})]\delta(\omega-\omega_{\xi})+n(\omega_{\xi})\delta(\omega+\omega_{\xi})
%\end{equation}
where $n(\omega)=(\exp({\omega/k_{\rm B}T)-1})^{-1}$ is Bose-Einstein distribution function.

The Auger autoionization, which for L-edge states of early transition metals is known to dominate the population decay,~\cite{Stoehr_book_1992} is incorporated  phenomenologically by the decay rate $\Gamma_a$  yielding the simple Auger decay matrix 
\begin{equation}\label{eq:Auger}
\mathcal{A}_{ab,cd} =-\delta_{ab}\delta_{cd}\delta_{ac} \Gamma_a \, .
\end{equation}
Notice that this term is not norm-conserving.
\begin{table}
\caption{Pulse train characteristics: wavelength of the driving laser $\lambda$, delay between two pulses in a sequence $t_i-t_{i-1}$ (see Eq.~\ref{eq:pulse}), duration of a single pulse $\sigma = T/14$ and $\sigma = T/28$, where $T$ is the period of the optical cycle of the driving laser
\label{tab:pulse}}
\begin{tabular}{p{1.5cm} p{2cm} p{1.5cm} p{1.5cm}}
$\lambda$, nm & $t_i-t_{i-1}$, fs & $T/14$, fs & $T/28$, fs \\
\hline
\hline
400 & 0.67 & 0.095 & 0.048 \\
800 & 1.33 & 0.191 & 0.095 \\ 
1400 & 2.34 & 0.334 & 0.167 \\
2000 & 3.34 & 0.476 & 0.238 \\
\hline
\end{tabular}
\end{table}

%--------------------------------------------------------------
\section{Computational details}
\label{sec:comp}
%--------------------------------------------------------------
The approach is applied to the spin-dynamics in the core-excited \Fe complex 
%(see Fig.~\ref{fig:system}a)) 
representing a model of the solvated Fe$^{2+}$ ion, see Fig.~\ref{fig:multi_pulse_scheme}b).
Its X-ray absorption and resonant inelastic X-ray scattering spectra were discussed in Refs.~\citenum{Bokarev_PRL_2013,Atak_JPCB_2013,Golnak_SR_2016}.
The ground electronic state of \Fe corresponds to the quintet ($S=2$) high-spin d$^6$ electronic configuration, which should be triply degenerate if octahedral symmetry is assumed.
This degeneracy is lifted due to the weak Jahn-Teller effect leading to the three close-lying electronic states.

CSFs were constructed within a RASSCF (restricted active space self-consistent field) scheme using an active space containing 12 electrons distributed over the three 2p (1 hole is allowed) and five 3d (full CI) orbitals. 
%(cf. Fig.~\ref{fig:system}b)) 
This setup is sufficient to describe the core excited electronic states corresponding to the dipole allowed 2p$\rightarrow$3d transitions.~\cite{Bokarev_PRL_2013,Atak_JPCB_2013,Golnak_SR_2016} 
%The number of holes in RAS1 was limited to one, whereas full CI was done within RAS2.  This active space included up to 4h4p configurations and resulted in 
In total, 35 quintet ($S=2$) and 195 triplet ($S=1$) electronic states, directly interacting via SOC according to the $\Delta S=0,\pm1$ selection rule, were considered.  
%Note that septet electronic states ($S=3$) are not possible with this active space\rt{, since 2p$\rightarrow$3d excitations cannot increase the number of unpaired electrons}. 
Accounting for the different $M_S$ microstates, the total number of  SF and SOC  states was 760, with 160 being valence and 600 core ones. 
%The respective calculations are denoted as RASCI(1,2) below. Notice that both, the account for  4h4p excitations and SOC are essential to recover the  dynamics of the highly correlated  core-excited states. In addition singlet states were also included  to test the influence of second-order SOC effects; denoted as RASCI(0,1,2). Their number increased the dimensionality of the basis by 170 states in total (50 valence and 120 core).  \rt{The atomic mean fields were generated by the [Ar]$4s^2 3d^6$ and [He]$2s^2 2p^4$ occupations for Fe and O neutral atoms, respectively.} 
Scalar relativistic effects we considered at the second-order Douglas-Kroll-Hess transformation level.~\cite{Douglas_AP_1974}  
%To correct for weak correlation effects the single-state second-order perturbation theory correction (RASPT2)~\cite{Malmqvist_2008} was  added to the diagonal of $\mathbf{H}_{\rm CI}$ matrix written in the SF basis, which was then back-transformed to the CSF basis. Thus, due to their diagonal nature this corrections influenced SOC only implicitly via the relative energetics of the interacting states. To avoid intruder states in RASPT2 calculations, an imaginary level shift~\cite{Forsberg_1997} of 0.4 E$_h$ was introduced. 1s,2s, 3s, and 3p orbitals of iron as well as 1s orbitals of oxygen were kept frozen in RASPT2.
Time-independent matrix elements in Eq.~\eqref{eq:Ham} were evaluated with the MOLCAS 8.0~\cite{Aquilante_JCC_2016} program package, applying the ANO-RCC basis set of triple-$\zeta$ quality.~\cite{Roos_JPCA_2004,Roos_JPCA_2005}

The electron-vibrational couplings in the SF basis, $g_{0n,\xi}$, were obtained from the forces at vertical excitation employing a quantum chemical hybrid approach on the basis of the shifted harmonic oscillators model.
Here, the force constant matrix was determined at the level of DFT/B3LYP/6-311G(d) for the ground state and the gradients in the excited electronic states were computed at the ground state geometry using the  RASSCF method as described above. The width $\gamma$ in the spectral density, Eq.~\eqref{eq:J}, accounting for the interaction between the modes of the solvated ion and further  degrees of freedom of the environment, was set to 500~cm$^{-1}$ for all normal modes. The  rates $\Gamma_a$, phenomenologically accounting for Auger decay, were set to 0.4 and 1.04\,eV for the L$_3$ and L$_2$ edges, respectively.~\cite{Ohno_JESRP_2009} These values correspond to lifetimes of the core hole of 10.3 and 3.98\,fs.

Equation~\eqref{eq:LvN} was solved using the Runge-Kutta-Fehlberg integrator with adaptive step size varying from 2.5\,as down to 0.09\,as depending on the field strength. The initial density matrix was populated according to the Boltzmann distribution at 300~K: 
\begin{equation}\label{eq:rho0}
\rho_{ab}(0)=\delta_{ab} \exp (-E_a/kT)/\sum_c \exp (-E_c/kT) \, .
\end{equation}
%In fact, the three lowest quintet states have non-negligible population, which yields  15 $M_S$ microstates. 

%The dynamics discussed here and first described in Refs.~\citenum{Wang_PRL_2017,Wang_MP_2017} is driven solely by the electronic coupling, while nuclei are fixed at the ground state equilibrium positions.\rt{~\cite{pierloot_relative_2006}} 
%\bt{We consider the clamped nuclei approximation to be valid since the few femtosecond dynamics modeled here is faster than the relevant vibrational periods of about 100~fs, see Ref.~\citenum{Wang_MP_2017} and discussion below.}

%The choice of the excitation regimes and pulse parameters in Eq.~\eqref{eq:pulse} is discussed in Section \ref{sec:results}.
%Due to the weak Jahn-Teller distortion the degeneracy is lifted resulting in three closely lying electronic states. 

The parameters $E_0$, $t_i$ and $\sigma$ in Eq.~\eqref{eq:pulse}, determining the shape of the time-dependent external electric field, were chosen to roughly resemble the regimes of the commonly used driving laser systems for the generation of high harmonics.
However, some  assumptions on the form of the incoming electric field have been done allowing to simplify the computational protocol.
For the wavelength of the driving laser, we have chosen 800\,nm corresponding to the Ti:Sapphire laser as well as  400,  1400, and 2000\,nm, which can be obtained by frequency doubling and parametric amplifiers of the 800\,nm laser. 
These values correspond to shifts between consecutive sub-pulses $t_i-t_{i-1}$ from about 1 to 7\,fs, see Table~\ref{tab:pulse}. Ten pulses have been considered in each series assuming equal envelopes of the subpulses for simplicity.
The simulations have been carried out for two field amplitudes, $E_0$, of 2.5 and 0.25\,a.u. for each pulse comb.
Note that despite the large values of $E_0$, at soft X-ray wavelength this corresponds to the weak field regime, see Refs.~\citenum{Wang_PRL_2017, Wang_MP_2017}.
The polarization vector $\vec{e}$ has been chosen to be parallel to the 
%\rt{2.0196\AA} 
shortest Fe--O bond of the \Fe complex.
It should be noted that the trains of ultrashort pulses in our theoretical study have been designed solely for illustration purposes and correspond to realistic experimental setups only approximately.
However to the best of our knowledge, such a setup is not yet available.
The real carrier frequencies used have not been set to particular harmonics of the driving laser and for convenience corresponded to maxima of two peaks in absorption spectrum (708.4 and 711.5\,eV, cf. Fig.~\ref{fig:multi_pulse_scheme}b)), considered before, see Refs.~\citenum{Wang_PRL_2017,Wang_MP_2017}. Fig.~\ref{fig:multi_pulse_scheme}b) also contains the HHG frequency combs generated with different lasers overlaid with the spectrum.
One can see that the chosen ``resonant'' carrier frequencies only slightly differ from the exact positions of the harmonics apart from the case of 400\,nm driving laser.
The widths $\sigma$ in Eq.~\eqref{eq:pulse} have been chosen as $T/14$ and $T/28$, where $T$ is the period of optical cycle of the driving laser. 
Here we assume that the parameter $\sigma$ can be controlled filtering out the low-energy harmonics of the HHG output.
(Otherwise, it would be bound to the inverse of the cutoff energy.)
The pulse characteristics are summarized in Table~\ref{tab:pulse}.
%
%--------------------------------------------------------------
\section{Results and discussion}
\label{sec:results}
\subsection{Role of electron-vibrational coupling}
\label{sec:bath}
%--------------------------------------------------------------
%
\begin{figure*}
\includegraphics[width=0.95\textwidth]{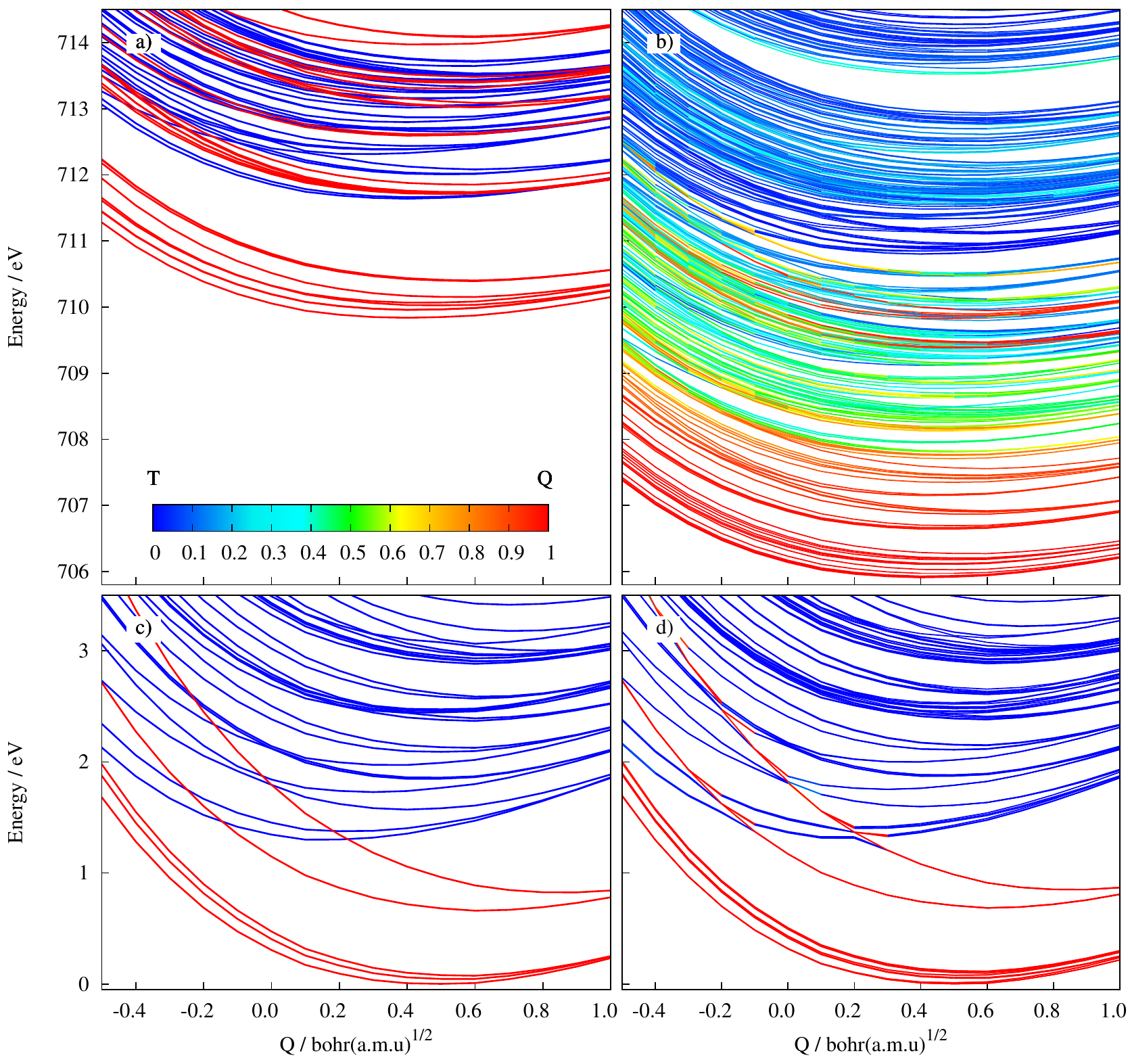}
\caption{The cuts of the potential energy surfaces along the most active tuning vibrational mode  (ground state frequency 417\,cm$^{-1}$). Panels a) and c): core and valence PESs of quintet SF states (red lines) and triplet SF states (blue lines), respectively, in the SF basis. Panels b) and d): same PESs in SOC basis, the color of lines corresponds to the collective contribution $\sum_{n,M_S} |C^{(S=2,M_S)}_{an}|^2$ of all quintet ($S=2$) SF states to a particular $\ket{\Psi^{\rm SOC}_{a}}$, see Eq.~\eqref{eq:SOC_states}. Note that only a part of states is shown in panels a) and b), corresponding mainly to the L$_3$ absorption band. Also note that due to the hybrid approach (DFT ground state geometry and RASSCF PES) the minima of the ground state PES is not at zero.
\label{fig:PESs}
}
\end{figure*}

\begin{figure}
\includegraphics[width=0.5\textwidth]{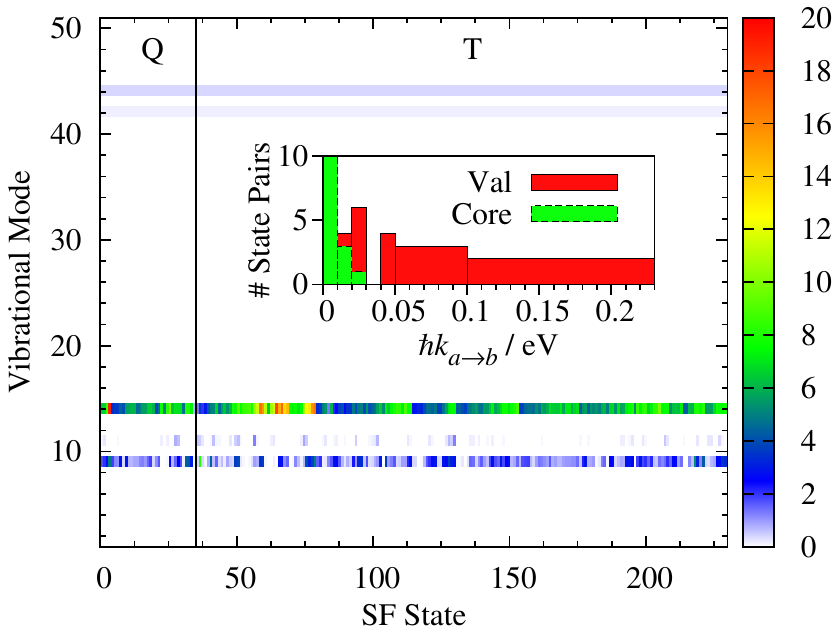}
\caption{Huang-Rhys factors, $g_{0n,\xi}^2/2$, for the transitions from the ground to the quintet (Q) and triplet (T) SF states for all 51 vibrational modes $\xi$.  Inset: Distribution of relaxation rate  $\hbar k_{a \rightarrow b}$ values (Eq.~\eqref{eq:k_ab}) for different pairs ($a,b$) of valence and core SOC states.  (Note that the first column of the histogram is cut at the value of 10.)
\label{fig:HR_matrix}
}
\end{figure}

In order to obtain an overview with respect to the effect of electron-vibrational coupling, we have calculated one-dimensional cuts of the potential energy surfaces (PES) along the symmetric Fe--O breathing mode ($\xi=15$), which is the  most strongly coupled mode (see below). Results are shown in Fig.~\ref{fig:PESs}, with the color code  indicating the change of multiplicity of the respective states from quintet (red) to triplet (blue). Panels a) and c) correspond to PES of core- and valence-excited electronic states, respectively, in the SF basis. The results for the SOC state representation are shown in panels b) and d). 

\begin{figure*}
\includegraphics{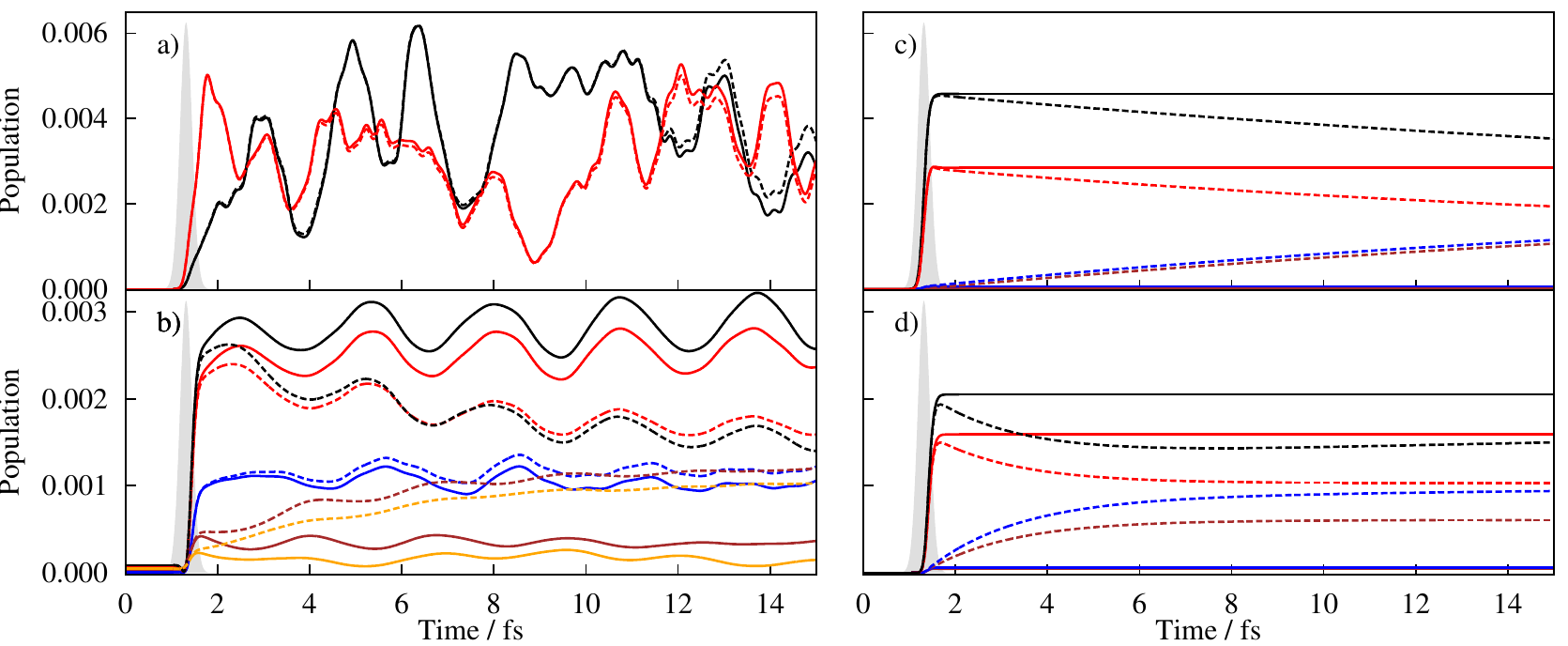}
\caption{Population of selected states corresponding to the largest relaxation rates (cf.~Fig.~\ref{fig:HR_matrix}) with (dash lines) and without  (solid lines) bath. a) core SF states; b) valence SF states; c) core SOC states; d) valence SOC states. The envelope of the excitation pulse is shown as a grey-shaded area.
\label{fig:pop_selected}
}
\end{figure*}

The states at lower energy are of quintet character in all cases. 
For the valence excited states this is followed by a band of triplet states, with some occasional mixing between triplet and quintet states in the SOC basis in the vicinity of intersection points. 
We also notice that triplet states cross each other. 
Since the selected mode is of tuning-type it can be anticipated that there are coupling modes, which render these crossings into conical intersections. 
Overall, the treatment of the dynamics in the valence excited states would require sophisticated wave packet propagation.

The situation is rather different  for the core-excited states, in particular, concerning the SOC state PES, which are of main interest here. 
Due to the large value of the SOC ($|V_{\rm SOC}|=8.5$\,eV) there are substantial energy shifts if panels a) and b) are compared. 
In addition, there is a splitting of the $M_S$-components, leading to an almost continuous density of states (especially in energy region 708-711\,eV which will be used for excitation below). 
In passing, we notice that in panel c) one can still observe a certain pattern with respect to the multiplicity of the states, which changes from quintet via strongly mixed to triplet with increasing energy. 

From this picture one can draw two conclusions: 
First, SOC is the dominant coupling mechanism, making electron-vibrational coupling a small perturbation at short times. 
This justifies the use of a Markovian Quantum Master Equation for the description of the effect of electron-vibrational coupling. 
Second, due to the essentially continuous density of states it can be expected that  details of the electron-vibrational interaction will only be of minor importance. 
Although the SF PES are manifestly anharmonic, the harmonic approximation in conjunction with the Huang-Rhys model might still give a reasonable estimate for the electron-vibrational coupling constants.    In fact, one might view the present approach as providing a simple kinetic approach to  the influence of electron-vibrational coupling on the spin-flip dynamics.

%\rt{First discuss potential curves in diabatic and adiabatic picture. The argument is that the density of states is high and the SOC coupling substantially changes the picture. That is why it is more corresponding to the incoherent dynamics with almost continuous spectral density rather than a coherent wavepacket dynamics along some normal modes. The same argument about quasi-continuous DOS explains Markov approximation since in this case the dephasing responsible for decay bath autocorrelation function is very fast.}
%In principle, nuclear motion can substantially influence electronic dynamics causing fast decoherence and the width of nuclear distribution can be the reason for ultrafast dephasing~\rt{[refs on dephasing]}.
%That is why the implicit inclusion of nuclear degrees of freedom via spectral density of the vibrational bath needs justification.
%To do so, we analyzed Potential Energy Surfaces (PESs) of valence and core excited states along normal modes of the \Fe system.
%Since in our case the heat bath represents a collection of harmonic oscillators it is convenient to perform the analysis in terms of shifted harmonic oscillators looking at so-called Huang-Rhys factors, see Section~\ref{sec:dissipation}.

%The calculations have been performed at a fixed nuclear geometry
Analysis of Huang-Rhys factors in the SF basis, $g_{0n,\xi}^2/2$, shows that only few vibrational modes out of the 51 possible ones are active, see Fig.~\ref{fig:HR_matrix}. Most prominent shifts between harmonic oscillators for both quintet and triplet as well as valence- and core-excited states correspond to two modes ($\xi=10$ and  15) with  ground state frequencies of 228 and 417~cm$^{-1}$, which are of Fe--O stretching type. This is not surprising since only oxygen 2p orbitals are strongly mixing with 3d orbitals of iron and thus both 3d-3d and 2p-3d electronic transitions are influenced by the distance to oxygens and are almost insensitive to vibrations of water ligands themselves. Nevertheless, some other vibrations like symmetric O--H stretching modes at 3640~cm$^{-1}$ ($\xi=43$) and 3645~cm$^{-1}$ ($\xi=45$) also have non-vanishing magnitudes, but do not play a major role.

%\rt{Some comments with this respect?
%\begin{itemize}
% \item crossings in the FC region
% \item could cause unltrafast nuclear driven ISC
% \item how relevant this should be if SOC is that strong
% \item discuss applicability of perturbation theory in principle
%\end{itemize}
%}
The rates calculated according to Eq.~\eqref{eq:k_ab} are summarized in a histogram shown as an inset in Fig.~\ref{fig:HR_matrix}.
Out of 288420 possible state pairs and thus transition rates, only few are larger than 0.01\,eV.
Remarkably, energy dissipation due to transitions between valence states is more prominent than that between core excited states.

The population dynamics for the selected valence and core SF and SOC states is presented in Fig.~\ref{fig:pop_selected}.
These states have been selected to show the effect of the largest  dissipation rates.
To simplify the analysis, the regime of excitation corresponds to the single 132\,as pulse, resembling that used in Ref.~\citenum{Wang_PRL_2017}, and neglecting Auger decay.
Due to the spectral bandwidth of ultrashort pulse many core states are excited.
The typical populations of individual states are comparable to that shown in Fig.~\ref{fig:pop_selected}a.
The populations of valence SF states show oscillations typical  for zero-order states building up a superposition in case of small coupling between them.
The core SF states in turn exhibit a more intricate pattern which can be explained by much stronger coupling and larger state density.
The eigenstates of the SOC Hamiltonian are stationary as soon as the pulse is over and no dissipation is accounted for.
Interaction with the vibrational bath causes notable population transfer between state pairs with largest $k_{a \rightarrow b}$.
However, due to the small amount of such pairs with $\hbar k_{a \rightarrow b} > 0.01$\,eV which are essentially responsible for energy dissipation to the vibrational bath, such a relaxation does not play decisive role at least at the timescale of 15\,fs which is very small if compared to typical period of Fe--O stretching vibrations ($>100$\,fs).
Although the populations of the particular states with the largest $k_{a \rightarrow b}$ are notably affected by the dissipation as is illustrated in Fig.~\ref{fig:pop_selected}, 
%again with the effect for valence states being more pronounced than for the core ones, 
the populations of these states themselves are quite small what does not lead to notable differences in the dynamics of the total populations of quintet 
%($\sum_{n,M_S} |C^{(S=2,M_S)}_{nk}|^2$) 
and triplet 
%($\sum_{n,M_S} |C^{(S=1,M_S)}_{nk}|^2$) 
states (compare magnitudes of populations with Fig.~\ref{fig:dynamics1}a).

Summarizing, within our model the SOC driving force (which is electronic in its nature) of the ultrafast spin flip is by far dominating over the effect of electron-vibrational coupling.  Essentially, the population relaxation due to vibrations can be neglected especially taking into account the ultrashort time period considered. Importantly, the dephasing caused by coupling to the nuclear bath is not destroying the coherence initially created by the absorption of the ultrashort pulse. The last statement, however, requires a note of caution. In our study, we did not account for the width of nuclear wave packet in the initial state which can also cause ultrafast dephasing.~\cite{Vacher_PRA_2015,Arnold_PRA_2017,Vacher_PRL_2017}%, see \Supp.

%--------------------------------------------------------------
\subsection{Multipulse excitation}
\label{sec:mult_pulse}
%--------------------------------------------------------------
%
\begin{figure*}
\includegraphics[width=0.9\textwidth]{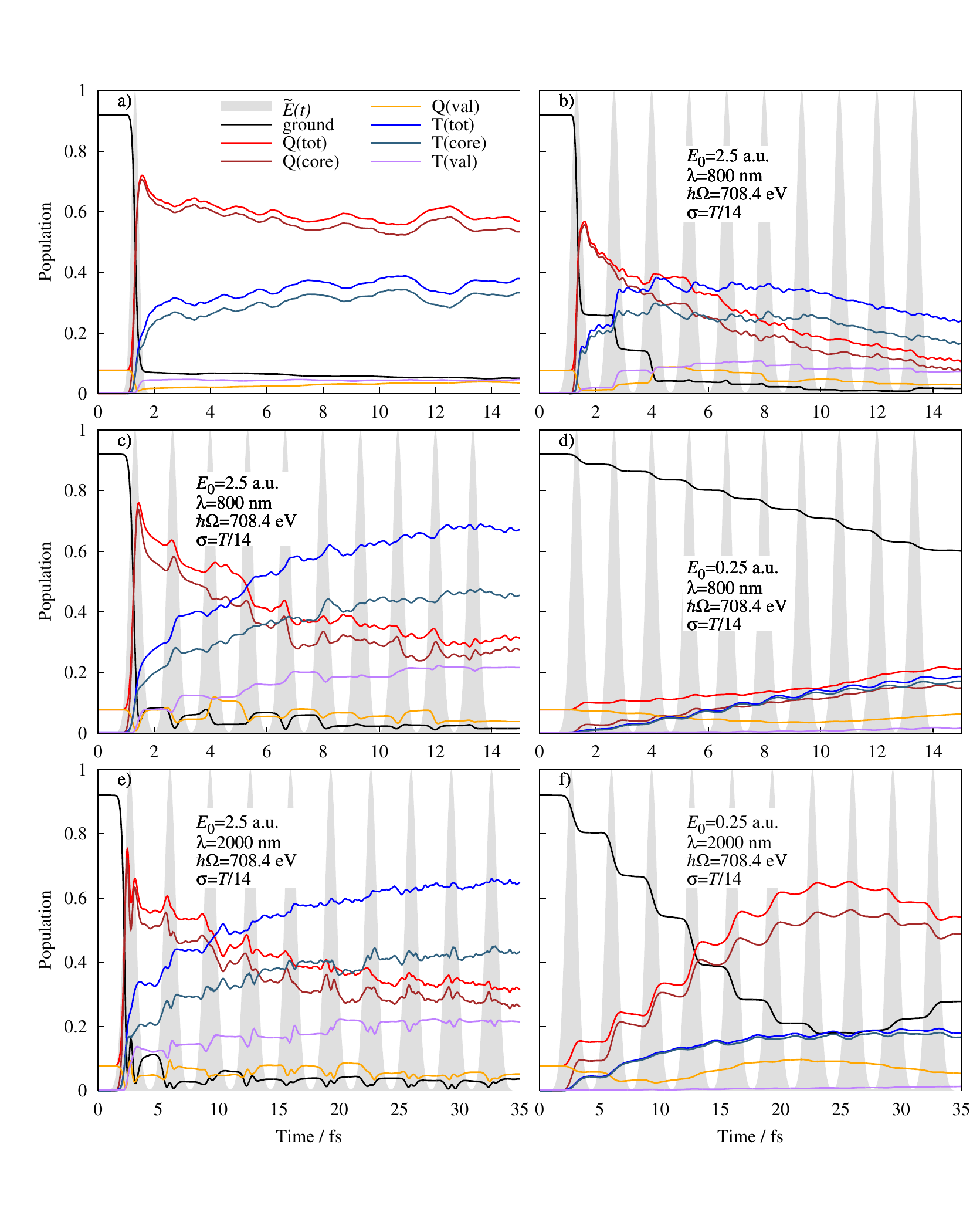}
\caption{Spin dynamics initiated by pulse trains %($\vec{E}(t)=\vec{e}\tilde{E}(t)\cos(\Omega t)$, see Eq.~\ref{eq:pulse}) 
having different characteristics: The shape of the normalized pulse envelope $\tilde{E}(t)/E_0=\sum_i\exp(-(t-t_i)^2/(2\sigma^2))$ is depicted with grey filled curves. The  populations of electronic SF states are presented for the cases of total quintet, Q(tot), and triplet, T(tot), as well as of the respective valence, Q(val) and T(val), and core, Q(core) and T(core). The black curve corresponds to the population of the lowest quintet (ground) state; note the finite temperature of 300\,K resulting in an initial population of 0.92. The amplitude $E_0$, wavelength of the driving laser $\lambda$, carrier frequency (energy) $\hbar \Omega$, and subpulse duration $\sigma$ are given in the respective panels. 
%In panels b)-e), the single subpulse duration is $\sigma=T/14$, where $T$ is the period of the driving pulse optical cycle. 
The dissipation due to vibrational bath is included. 
Panel a): Dynamics initiated by a single pulse ($E_0=$2.5\,a.u., $\hbar \Omega=708.4$\,eV, $\sigma=$132\,as), no vibrational bath, no Auger decay. Panel b): Example of dynamics accounting for Auger decay. Panels: c)-f) Auger decay is excluded for  simplicity of the analysis.
%Single subpulses duration are \rt{191 (panels a) and b)) and 95\,as} corresponding to bandwidths $\hbar/\sigma=$ 21.7 and 43.5\,eV, respectively. 
%Field amplitudes are $E_0=$\rt{2.5}\,$\rm E_he^{-1}bohr^{-1}$ for panels a) and b) and 0.25\,$\rm E_he^{-1}bohr^{-1}$ for panels c) and d).
%Auger decay is not accounted for.
%Total populations of core- and valence-excited SF states with different multiplicity are shown as: ground quintet (blue line), quintet valence other than the ground one (light green), triplet valence (purple), triplet core (magenta). The envelope of the excitation pulse train is shown as filled grey curve. 
%The population of $M_S$-components of the ground and first two excited states is shown by the blue line.  
\label{fig:dynamics1}
}
\end{figure*}

\begin{figure*}
\includegraphics[width=0.9\textwidth]{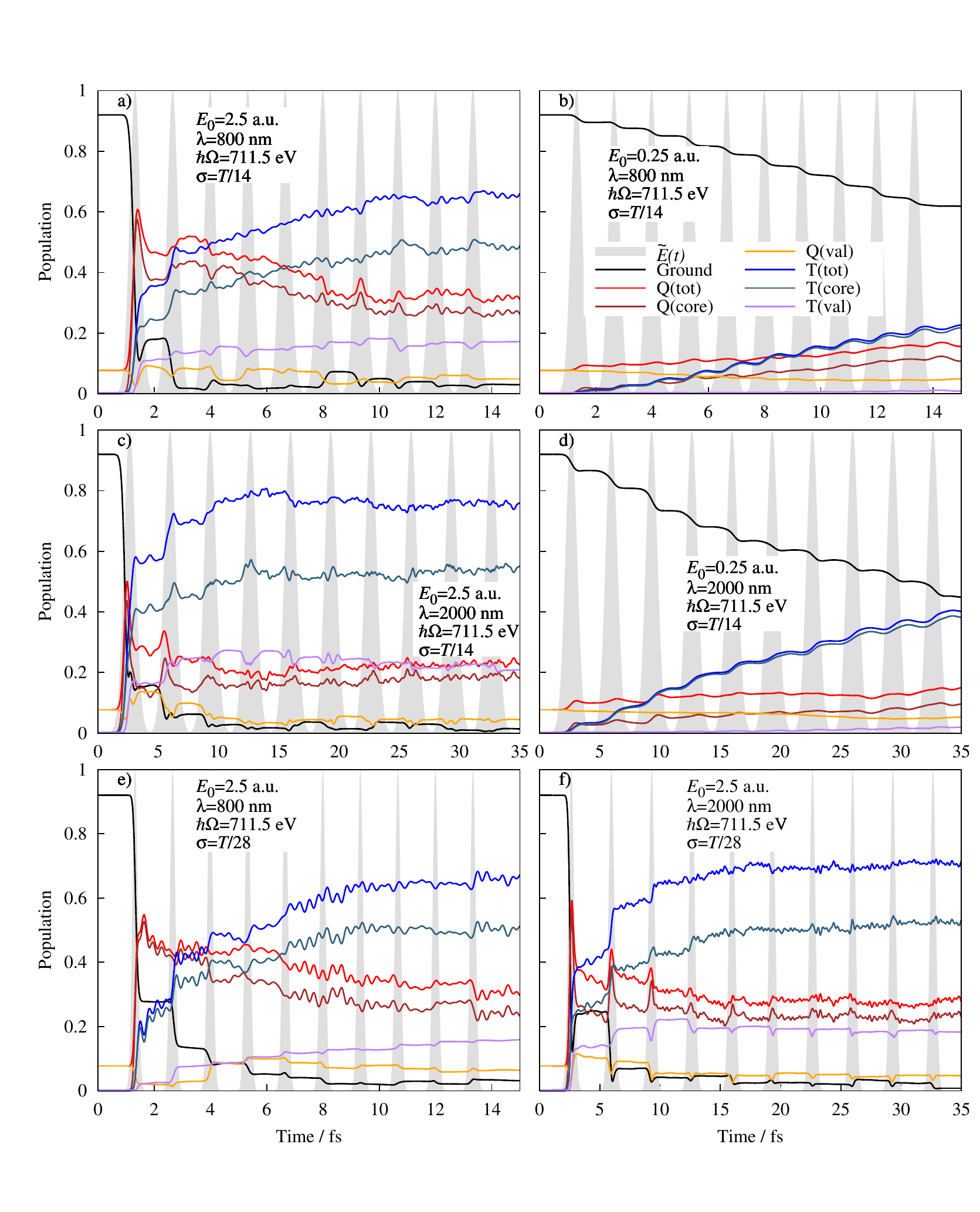}
\caption{Spin dynamics initiated by pulse trains %($\vec{E}(t)=\vec{e}\tilde{E}(t)\cos(\Omega t)$, see Eq.~\ref{eq:pulse}) 
having different characteristics as given in the panels; see also Fig.~\ref{fig:dynamics1}. 
\label{fig:dynamics2}
}
\end{figure*}

\begin{figure}
\includegraphics[width=0.47\textwidth]{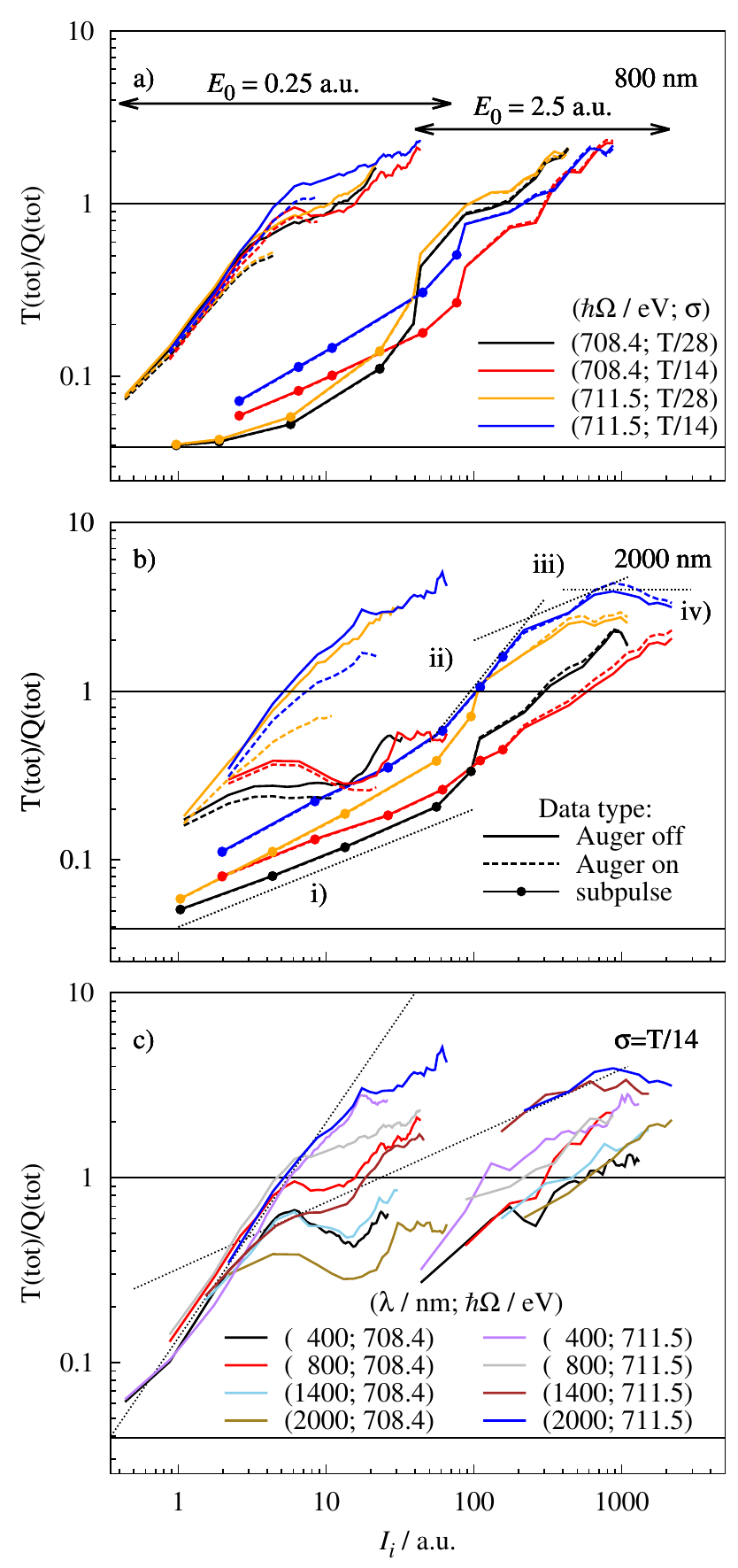}
\caption{
Relative yield of the total triplet population, T(tot), with respect to the total quintet one, Q(tot), versus the integrated intensity envelope $I_i=\int_{-\infty}^{(t_{i+1}-t_i)/2} |\tilde{E}(t)|^2 dt$. Here, the time $(t_{i+1}-t_i)/2$ corresponds to end of the $i$th subpulse in a train before $i+1$th  subpulse starts. Panels a) and b) correspond to  $\lambda=800$ and 2000\,nm. The population dynamics with/without Auger decay are shown as solid/dashed lines, respectively. 
%\bt{All panels contain two groups of curves with $E_0=0.25$ and 2.5\,a.u. which are barely overlapping along the intensity axis.} 
\bt{
%To increase the overlap between the two branches, 
%The first subpulse of the sequence with 
For $E_0=2.5$\,a.u. in panels a) and b), the first subpulse in the sequence is split into parts to ensure the overlap of two groups of curves along the $I_i$-axis, see text.} The respective data are shown with filled circles. In panel c), the results for different $\lambda$ are presented for $\sigma=T/14$. Two horizontal lines denote the starting T(tot)/Q(tot) and the point of T(tot)=Q(tot).
\label{fig:integrated_intensity}
}
\end{figure}

It has been shown previously that a single intense soft X-ray pulse can cause an unprecedented ultrafast spin-flip.~\cite{Wang_PRL_2017, Wang_MP_2017}
An illustration for this process is given in Fig.~\ref{fig:dynamics1}a.
Here, the total populations (Q(tot) and T(tot)) of quintet, $\sum_{n} \rho_{nn}^{(S=2)}$, and triplet, $\sum_{n} \rho_{nn}^{(S=1)}$,  SF states are shown as a function of time.
In addition, the overall populations of valence and core quintet and triplet states are given, where summation over $n$ is performed only within the respective manifolds.
One can see that an isolated pulse with duration 132\,as initially excites the system to the quintet core manifold and subsequently the population transfers to the triplet manifold on the timescale of few fs due to strong SOC. This process is purely electronically driven and occurs even  when the nuclei are fixed and no coupling to the vibrational bath  is accounted for.~\cite{Wang_PRL_2017, Wang_MP_2017}

Since isolated few-femtosecond and sub-femtosecond pulses 
%are to date quite exotic and 
require special techniques to be obtained,~\cite{Schultz_book_2014} we have investigated what happens if pulse trains are applied. The parameter space $(\lambda, \sigma, E_0, \hbar \Omega)$ of these pulse bursts (see Section~\ref{sec:comp}) has been systematically screened.
%A number of parameters has been chosen as described in Section~\ref{sec:comp}.
In the following, we discuss the main trends on the basis of few illustrative examples shown in Figs.~\ref{fig:dynamics1}b)-f) and~\ref{fig:dynamics2}; more  results can be found in the \Supp. 
Both 
%Fig.~\ref{fig:dynamics1}} 
figures focus on two driving laser wavelengths, i.e.\ 800\,nm and 2000\,nm.
The first one corresponds to the most common Ti:saphire laser, typically used for HHG experiments, whereas the second is available via parametric amplifiers and should allow for 
%be more feasible experimentally due to 
higher cutoff energies possibly reaching soft X-ray range with higher intensity. 
Note  that $\lambda$ actually fixes the interpulse delay to the half of period. 
Moreover, in our model, the duration of individual subpulses has been chosen to depend on the period ($\sigma=T/14$ and $T/28$) and thus $\lambda$ itself, see Sec.~\ref{sec:comp}. 
This makes the wavelength of the driving laser the most important parameter in our simplified model. 
Further, to address the field strength dependence, the results are presented for $E_0=2.5$\,a.u. and 0.25\,a.u. with respective intensities 8.8$\times$10$^{16}$ and 8.8$\times$10$^{15}$ W/cm$^2$. 
The carrier frequencies of $\hbar\Omega=$708.4 (Fig.~\ref{fig:dynamics1}) and 711.5\,eV (Fig.~\ref{fig:dynamics2}) correspond to the maxima of two peaks in L$_3$-edge absorption spectrum of aqueous Fe$^{2+}$ ion, see Fig.~\ref{fig:multi_pulse_scheme}b).~\cite{Golnak_SR_2016} 
These regions resemble  different spin-orbit coupling regimes, see for instance Ref.~\citenum{Wang_PRL_2017}. At 708.4\,eV the prepared superposition of states contains more quintet SF states, whereas at 711.5\,eV more triplet ones as is apparent from Fig.~\ref{fig:PESs}b).
%correspond roughly correspond to the centers of the Gaussians in the frequency domain at energies where SF states are mixed differently and thus resemble slightly different SOC regime, see Refs.~\citenum{Wang_PRL_2017, Wang_MP_2017}.
%This argument can be understood from Fig.~\ref{fig:PESs}b), where the center of the pulse lineshape at 708.4\,eV corresponds to the mostly quintet states with minor admixture of triplets, whereas for 711.5\,eV it represents an intricate mixture of SF states of both multiplicities.
%The change of the carrier frequency has been shown to be notably changing the dynamics for the case of isolated pulses~\cite{Wang_PRL_2017, Wang_MP_2017} and is also studied here for bursts of pulses.

%\rt{Explain different kinds of curves in these figures.}

Looking at the results in Figs.~\ref{fig:dynamics1}c)-f) and~\ref{fig:dynamics2} one can see that the pulse trains essentially demonstrate similar effects on the spin dynamics of the system as the isolated pulses (cf. 
%case II in Refs.~\citenum{Wang_PRL_2017, Wang_MP_2017} and 
Fig.~\ref{fig:dynamics1}a), but with a number of differences. 
First, there are prominent ultrafast dynamics happening between individual pulses in a train. 
It is most obvious in case of high field intensities. % presented in panels \rt{c) and e)}. 
These dynamics corresponds to the field-free evolution of the coherent electronic wave packet created by a single pulse. As a result the population of SF quintet states drops down and that of SF triplet states increases between pulses. 
Note the close similarity between the initial region of the population curves, e.g., in panels a) and c) of Fig.~\ref{fig:dynamics1}. 

Second, a remarkable feature of the dynamics is the stepwise pumping of the triplet population with each new incoming subpulse if $E_0=$2.5\,a.u. is used. 
%The first step, being the largest, can be up to almost 0.4 high. 
In general, the height of the steps is almost monotonously decreasing with every next pulse and the population approaches a plateau, which only slightly depends on the pulse characteristics and in most cases amounts to around 0.7.
%, with $\lambda=$400\,nm and $\hbar \Omega=$708.4\,eV being an exception, see \Supp.
For instance, for $\lambda=$800~nm and 2000~nm and $E_0=$2.5\,a.u. 
%\rt{(panels c) and e))},  
the first five or six pulses correspond to largest steps in triplet population growth and subsequent ones lead to relatively minor changes in population, comparable to those of free dynamics (panel~\ref{fig:dynamics1}a)). 
%\rt{For $\lambda=$1400~nm case, the plateau is reached after first three pulses, see \Supp.} 
Thus, every next subpulse increases the triplet yield, if only final populations are in focus.
Observing the stepwise behavior in experiment would, however, require a pump-probe setup, what makes such an observation challenging.

In some cases, valence excited states gain substantial population. 
For instance, valence triplet states constitute up to 35\% of the total triplet population. %\rt{(panels c) and e))}. 
This effect may be clearly seen in the changes of populations during pulses with $E_0=$2.5\,a.u., where both triplet and quintet valence populations may rise during the pulse, and may be attributed to stimulated emission. However, this effect may be considered as being of limited practical relevance, since it requires very high field strengths.
%the field strength is too high and has been selected only for illustration purposes.

The character of the dynamics changes quite notably when the strength of the pulses is decreased by a factor of ten (Fig.~\ref{fig:dynamics1}d) and f) as well as Fig.~\ref{fig:dynamics2}b) and d)).  In this case, the contribution from stimulated emission is barely seen, apart from panel~\ref{fig:dynamics1}f).
%(apart from $\lambda=$2000\,nm and $\hbar \Omega=$708.4\,eV). 
The behavior of the system shows a much more regular stepwise pumping with the sizes of the steps being almost constant and much smaller than for the stronger fields. 
%The differences between strong and weak pulses are more pronounced than differences between different bursts parameters of strong pulses. 
%In particular, 
The yield of triplet states in most cases stays below that of quintet states, with the examples shown in panels \ref{fig:dynamics1}b) and \ref{fig:dynamics1}d) being rather an exception (see, e.g., Fig.~\ref{fig:integrated_intensity}). 
The qualitative behavior of quintet and triplet populations with respect to each other is not changed when Auger decay is included apart from the double-exponential total decay (see Fig.~\ref{fig:dynamics1}b as well as \Supp).

One can also analyze the dynamics in terms of the integral intensity (energy) transferred from the pulse to the system.
For this purpose, we have plotted in Fig.~\ref{fig:integrated_intensity} the total triplet yield relative to the total quintet population, T(tot)/Q(tot), versus the integrated intensity envelope $I_i=\int_{-\infty}^{(t_{i+1}-t_i)/2} |\tilde{E}(t)|^2 dt$.
Here, $i$ runs over individual pulses in a train and, thus, the triplet yield after every subpulse is plotted.
\bt{The data represent two groups of curves for $E_0=0.25$ and 2.5\,a.u. which are barely overlapping along the intensity axis if integer number of pulses is considered. 
To compare the triplet yield for both $E_0$ at the same value of integrated intensity,  
%increase this overlap 
we have split the first pulse in the sequences with $E_0=2.5$\,a.u. into pieces of shorter duration (and thus smaller intensity).}  
%and $\lambda=800$ (panel a)) and 2000\,nm (panel b)).}
The respective data points are depicted by the filled circles in Fig.~\ref{fig:integrated_intensity}a) and b).
The data with Auger decay taken into account are plotted as dashed lines; these data sets included always 10 pulses in a train.
For dynamics without Auger effect and 0.25\,a.u. field amplitude, longer pulse sequences are applied again to increase \bt{the overlap in integrated intensity} with stronger pulses. 
Further, results for different $\lambda$ and $\hbar \Omega$ are given in panel c) for $\sigma=T/14$.

The dependencies in Fig.~\ref{fig:integrated_intensity} can be roughly subdivided into several almost linear regions (see Fig.~\ref{fig:integrated_intensity}b); note the double logarithmic scale): i) initial one, mainly seen during the first pulse for $E_0=2.5$\,a.u.; ii) linear rise with high steepness most apparent for $E_0=0.25$\,a.u; iii) rise with lower steepness; iv) saturation region which can be followed by decrease or oscillatory behavior of the T/Q ratio.
The region iii) with lower steepness begins when the triplet population starts to dominate the quintet one. 
That is why it can be attributed to the interplay of the absorption/emission and Q$\rightarrow$T/T$\rightarrow$Q forward and backward processes, thus, decreasing the rate of the spin crossover.
Remarkably, the steepness of the linear regions i)-iii) is quite similar for different pulse characteristics, see, e.g., Fig.~\ref{fig:integrated_intensity}c).
Similarity of all dependencies underlines the statement that the essential dynamics does not crucially depend on the pulse strength what justifies selection of high intensities for illustration purposes.

Pulse duration or associated bandwidth has also an influence on the dynamics. %as is illustrated in the \Supp.} 
In general, shorter durations ($\sigma=T/28$) cause faster oscillations than the longer ones ($\sigma=T/14$), compare \ref{fig:dynamics2}a) and \ref{fig:dynamics2}c) with \ref{fig:dynamics2}e) and \ref{fig:dynamics2}f). 
This can be rationalized by the larger bandwidth and thus involvement of more energetically distant states in the superposition. 
Naturally, since $E_0$ is the same for both durations, shorter pulses have smaller area under the $|\tilde{E}(t)|^2$ curve and thus are effectively weaker than the longer ones (note the shift along $x$-axis in Fig.~\ref{fig:integrated_intensity}).

Interestingly, pulse-train excitation with weaker fields shows more variability than with stronger ones. 
This applies both to time-dependences of populations (Figs.~\ref{fig:dynamics1} and~\ref{fig:dynamics2}) and integral intensity picture (Fig.~\ref{fig:integrated_intensity}).
For weaker pulses, the system is not kicked that hardly and evolution is closer to its natural dynamics. This is not the case if trains of pulses with $E_0$=2.5\,a.u. are used. Here, populations of all kinds of states show a similar pattern across the pulse parameter space.
Importantly, at both carrier frequencies triplet population exceeds the quintet one reaching a plateau at similar values.
%\rt{The bandwidth of the incoming light is very broad and exact position of the Gaussian lineshape in the frequency domain is almost irrelevant.}
%in case of bursts of 10 pulses the population of triplet states in all cases (apart from \rt{...}, see Supplement) exceeded that of the total population of quintet states.
In contrast, weaker pulses ($E_0=$0.25\,a.u.) initiate the dynamics that are much more similar to those triggered by a single pulse.~\cite{Wang_PRL_2017,Wang_MP_2017}
For instance, two carrier frequencies ($\hbar \Omega=$708.4 and 711.5\,eV) and pulse durations ($\sigma=T/14$ and $\sigma=T/28$) correspond to different yields of triplet states. 
%In this respect, the dynamics driven by a single pulse, and namely quintet/triplet population ratio at 15\,fs after the pulse, has been previously found to be sensitive to the carrier frequency and width of the pulse.~\cite{Wang_PRL_2017,Wang_MP_2017} 
%Here, the coherent superpositions prepared at two carrier frequencies are more different and initiated dynamics are almost insensitive to the pulse characteristics in the strong pulse regime. 
However in general, one can say that dynamics driven by a series of pulses seems to be less dependent on the characteristics of a particular pulse in a train than in case of isolated pulses.

A possible explanation could be that a single pulse creates a coherent superposition which evolves in time according to the full Hamiltonian in Eq.~\eqref{eq:Ham}, while multiple pulses constantly prepare new superpositions smearing out the populations over core- and valence-excited states in course of absorption and stimulated emission.
%Every next pulse alters the natural dynamics initiated by an isolated subpulse.
This smearing over states with different energies makes the overall dynamics having less dependence on the position of the center and bandwidth of the incoming field envelope in the frequency domain. It could also be the reason for the plateau-like behavior of the triplet populations in case of the stronger field.
%carrier frequency ($\hbar \Omega$), single pulse duration ($\sigma$)

Surprisingly, the delay between consecutive pulses has also quite moderate influence on the dynamics. 
Only for %$t_i-t_{i-1}$=0.67\,fs 
$\lambda=$400\,nm the pattern of the fast oscillations notably changes. 
It corresponds to the shortest pulses used in this work meaning the largest bandwidth in the frequency domain. 
%In this respect, it is quite close to case/regime I in Refs.~\citenum{Wang_PRL_2017, Wang_MP_2017} and one sees ultrafast oscillations with a period of 0.32\,fs. 
The ultrafast oscillations with a period of 0.32\,fs can be then attributed to the superposition of the most distant states split due to SOC by 12.7\,eV.
%It can be particularly seen for the case of $\lambda=$400\,nm and $\hbar \Omega=$710\,eV where 

Finally, the leak of the norm due to Auger process is very fast (Fig.~\ref{fig:dynamics1}b)), corresponding to biexponential decay with time constants 3.98 and 10.34\,fs in accordance with our phenomenological model (Eq.~\eqref{eq:Auger}).
Despite of this fast decay, the spin-flip process is still faster leading to a behavior similar to that discussed above.
The only difference is that in this case the total triplet population is lower and is decaying with the slow Auger component.
In this respect, depositing population into valence triplet manifold enabled by stimulated emission decreases the destructive effect of the Auger decay on the spin-flip efficiency.
This scenario, however, would require very intense fields which are not yet achievable in practice.

The same conclusions can be drawn from Fig.~\ref{fig:integrated_intensity}.
There, the Auger decay naturally shows up for longer times (larger number of subpulses) and that is larger integrated intensities. 
At the beginning (region ii)) the population curves with and without Auger autoionization almost coincide and the divergence becomes more obvious for large intensities.
Interestingly, Auger decay has a major effect for sequences of weak pulse  ($E_0=0.25$\,a.u.), where it tends to notably decrease the triplet yield. 
For stronger fields, the effect is smaller and, in contrast, the triplet yield increases.

Summarizing, weak pulses are more selective. 
Further, for a particular integrated intensity without Auger decay, they are more efficient than the stronger ones if we compare triplet yields for a fixed $I_i$ (Fig.~\ref{fig:integrated_intensity}).
Accounting for the fact that they are more readily available in experiment, this makes them the most suiting candidates to address or steer ultrafast spin dynamics.
This effect is, however, counterbalanced by the destructive influence of Auger decay.
One the other hand, from the viewpoint of the time of the spin crossover, stronger pulses should be more efficient as they lead to faster increase of triplet population.
In this case, the Auger effect is less important because of faster spin-flip transitions and involvement of non-decaying valence states via stimulated emission.

%\rt{Mention that multiple pulses transfer population also to the triplet valence manifold. This should slow down Auger decay. Mention how Auger is influencing the dynamics.}

%\rt{For the weak pulses conclusions are more like for single pulse: changing carrier frequency changes the quintet to triplet ratio.}

\iffalse
\begin{figure*}
\includegraphics[width=0.95\textwidth]{./images/Fig_713_1400_VC.eps}
\caption{Another figure to include.  
\label{fig:dynamics_713_1400}
}
\end{figure*}
\fi

%--------------------------------------------------------------
\section{Conclusions}
\label{sec:concl}
%--------------------------------------------------------------
This study represents an extension of the  analysis of  ultrafast spin-flip dynamics reported before.~\cite{Wang_PRL_2017, Wang_MP_2017}  The main two questions addressed here are: Does the coupling of the electronic subsystem to nuclear vibrations alter the dynamics? What changes if pulse trains are used instead of a single pulse?

The former problem is treated at the level of system--bath partitioning.  This approach corresponds to an approximate scheme since no non-adiabatic coupling is taken into account explicitly. Moreover, the width of nuclear distribution in the ground state is not accounted for. The bath is represented essentially by a single damped totally-symmetric vibrational mode which is strongly coupled to the  electronic states.  The effect of energy dissipation and decoherence is taken into account using a simple Bloch model. The computed rates are very small in comparison to the strength of SOC.  Moreover, the state pairs which are notably coupled by the symmetric stretching mode are very small in number. 
Therefore, within our dissipative model, the vibrational bath does not play prominent role on dynamics at least at the considered time-scale below 50\,fs.

Further, the effect of the multi-pulse excitation typical for HHG setups has been simulated.  Parameters of the pulses have been chosen to roughly resemble possible experimental setups. 
The net effect of interaction with pulse trains is in general analogous to isolated pulse excitation. 
Most important is that ultrafast spin-flip also occurs for multiple pulses. 
Even more, utilizing bursts makes the process more efficient leading to a stepwise pumping of triplet population. 
Interestingly, the differences in dynamics caused by intense pulses show less sensitivity to individual pulse characteristics such as carrier frequency, duration, and interpulse delay. 
Weaker pulses are more selective and in some cases show no predominant population of spin-flipped states.

The effect, studied here theoretically, calls for experimental observation. 
Our study suggests that the light characteristics needed to trigger this effect are within reach.
We are confident that further development of non-linear X-ray spectroscopic techniques such as stimulated resonant inelastic X-ray scattering~\cite{Zhang_TCC_2015} or employment of magnetic circular dichroism probe within a pump-probe experiment will enable measuring ultrafast spin-flip dynamics.

Although it is of fundamental interest to address ultimate timescale of electron dynamics caused by spin-orbit coupling, we envisage  that this effect could  be used for clocking ultrafast events. 
In this respect, it is of core-hole clock type~\cite{Piancastelli_JPB_2014} but has a different nature. 
Moreover, in case of spin-flips the characteristic timescale may be varied by addressing different atoms in the system or various types of core holes as well varying the carrier frequency and bandwidth of the incoming radiation, thus adjusting the strength of the coupling and thereby determining essentially the measured time window.

\section*{Acknowledgments}
The authors would like to acknowledge the fruitful discussion with Prof. Dr. Stefan Lochbrunner (University of Rostock) as well as financial support from the following sources: Deanship of Scientific Research (DSR), King Abdulaziz University, Jeddah, grant No. D-003-435 (H.W. and O.K.), Landesgraduiertenf\"orderung of Mecklenburg-Vorpommern (T.M.), and Deutsche Forschungsgemeinschaft grant No. BO 4915/1-1 (S.I.B.).

%\bibliographystyle{apsrev4-1}
%\bibliography{Spin_dynamics_3}%,reference}


\end{document}